\documentclass[twocolumn,prl]{revtex4}% Physical Review B RC

\usepackage{graphicx}
\usepackage{dcolumn}
\usepackage{amsmath}

\begin{document}
\preprint{HEP/123-qed}

\title[Short Title]{Laser-Induced Above-Bandgap Transparency in GaAs}
\author{Ajit Srivastava}
\author{Rahul Srivastava}
\author{Jigang Wang}
\author{Junichiro Kono}
 \thanks{To whom correspondence should be addressed}
 \homepage{http://www.ece.rice.edu/~kono}
 \email{kono@rice.edu}
\affiliation{Department of Electrical and Computer Engineering,
Rice Quantum Institute,
and Center for Nanoscale Science and Technology,
Rice University, Houston, Texas 77005, USA}

\date{\today}

\begin{abstract}
We report the observation of large ($\sim 40\%$) laser-induced
above-bandgap transparency in GaAs at room temperature.  The induced
transparency is present only during the pulse width of the driving midinfrard laser
pulses and its spectral shape is consistent with a laser-induced blue shift
of the band edge.
Our simulations based on the dynamic Franz-Keldysh effect reproduce the
salient features of the experimental results, demonstrating in particular
that the amount of the band edge shift is approximately given
by the ponderomtive potential.
\end{abstract}
\maketitle

%%%%%%%%%%%%% Introduction: Laser-Induced Transparency %%%%%%%%%%%%%%%%%%%
Under certain conditions, a normally opaque medium can be made transparent
through coherent interaction with laser light.
Such laser-induced transparency phenomena include self-induced transparency
\cite{McCallHahn67PRL} and electromagnetically induced transparency
\cite{Harrisetal90PRL}, which
have been extensively studied in atomic and molecular systems.
%%%%% Laser-Induced Transparency in Solids -- Fascinating but Challenging %%%%%
Solids, semiconductors in particular, are an obvious alternative to gases for further
exploring such coherent light-matter interactions.  The direct interplay of
the spatially periodic lattice potential and the temporally periodic laser field
is predicted to cause drastic band structure modifications such as band gap
distortion and oscillations \cite{BalkareiEpshtein73SPSS,Miranda83SSC},
dynamic localization \cite{DunlapKenkre86PRB}, and
the appearance of new gaps \cite{Kaminski93APP}.
In addition, semiconductors exhibit various many-body effects
\cite{ChemlaShah01Nature} and hence become
an even richer system for such studies.  Furthermore, the effect of quantum
confinement is predicted to exhibit exotic quantum dynamics, including coherent
destruction of tunneling \cite{GrossmannetAl91PRL}, collapse of minibands
\cite{Holthaus}, and quantum chaos \cite{Holthaus2}.
However, these phenomena have been mostly unexplored experimentally
due to the high required laser intensities and unavoidable sample damage.

%%%%%%%%%%%%% Key: Intense Long-Wavelength Light %%%%%%%%%%%%%%%%%%%%%
%With the advent of ultrafast femtosecond lasers and optical parametric
%amplifiers (OPAs), high intensity mid-infrared (MIR) radiation which is
%needed to see such effects is achievable without the threat of sample damage.

%%%%%%%%%%%%%%%%%%%%%%% Here we report ... %%%%%%%%%%%%%%%%%%%%%%%%%%%%%
Here, we demonstrate a new type of laser-induced transparency effect
in semiconductors.  The key to this effect is the use of intense {\em midinfrared} (MIR)
light, which helps minimize interband absorption and sample damage while
%employing the
%low dispersion existing at longer wavelengths and
maximizing the ponderomotive
potential $U_p$ (or the ``wiggle'' energy) \cite{intenselasers,Chinetal00PRL},
which increases quadratically with increasing wavelength.  For a driving laser field
with vector potential $A = A_{0} e^{i\omega t}$,
\begin{eqnarray}
U_p = \frac{e^2A_{0}^{2}}{4mc^2} =
\left(\frac{2\pi e^2}{mc}\right)\left(\frac{I}{\omega^2}\right),
\label{Up}
\end{eqnarray}
where $I$ is the intensity of the laser field, $e$ is the electronic charge,
$c$ is the speed of light in vacuum, and $m$ is the electron mass.
When $U_p$ becomes comparable to, or larger than, a characteristic energy of the system
(such as the ionization potential), {\em extreme} nonlinear
optical effects are expected
\cite{BucksbaumetAl87JOSAB,ChangetAl97PRL,Chinetal01PRL}.
Under such conditions we observed ultrafast photoinduced
transparency right above the band edge of GaAs at room temperature.
The effect can be interpreted as a laser-induced blue shift of the band edge,
whose amount is given by $U_p$.
In addition, photo-induced absorption below the band edge with unusual
laser power dependence was observed.  We explain these observations
in terms of the dynamic Franz-Keldysh effect (DFKE)
\cite{Yacoby68PR,JauhoJohnsen96PRL,Nordstrometal}.
%and present a reasonable theoretical fit to our data.
%This observed photoinduced transparency, is
%to our knowledge the first experimental observation of blue-shift of the band edge
%due to the DFKE. It represents
%a novel phenomenon quite unlike the well known self-induced or
%electromagnetically induced transparency and (as we shall see
%later) can be explained only on the basis of the DFKE.

%%%%%%%%%%%%%%%%%%  FIG. 1  %%%%%%%%%%%%%%%%%%
\begin{figure}
\includegraphics [scale=0.55] {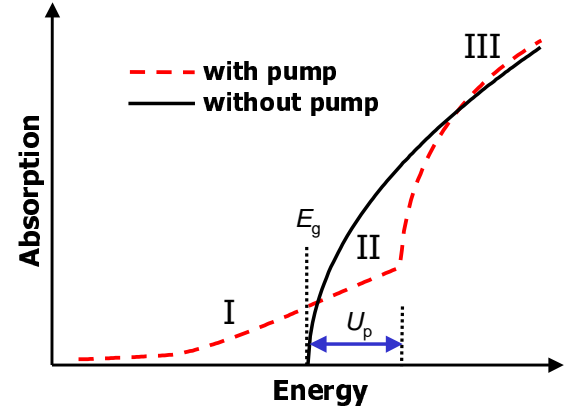}
\caption{Interband absorption in a typical direct-gap semiconductor near the band gap $E_g$
with (dashed line) and without (solid line) a strong driving field.  The dynamic
Franz-Keldysh effect predicts: (i)
below band gap absorption (region I), (ii) blue shift of the band edge causing an
induced transparency (region II), and (iii) oscillatory behavior above
the band gap (region III).  The magnitude of the blue-shift is predicted to be
equal to the ponderomotive potential $U_p$.}
\label{cartoon}
\end{figure}

%%%%%%%%%%%%%% What is DFKE? %%%%%%%%%%%%%%%%%%%
The DFKE is an intense-field effect in which changes are induced in the
transmission spectra near the fundamental band edge of a semiconductor in the
presence of an intense, high frequency field
\cite{Yacoby68PR,JauhoJohnsen96PRL,Nordstrometal}.
These changes include absorption below the band edge, a blue shift of the band
edge leading to induced transparency, and oscillatory behavior above the band edge
(see Fig.~\ref{cartoon}).
%, where the transition rate
% near the band edge of a typical bulk semiconductor is plotted.
Also, nonlinear mixing between the strong pump and a weak probe beam leads to
the generation of optical sidebands
\cite{Konoetal97PRL,Nordstrometal,side3,side4,Chinetal01PRL}.
The DFKE, which is a nonresonant (far from resonance) effect, is quite
different from the AC Stark effect which occurs in the
presence of strong driving fields resonant (or nearly resonant) with
electronic transitions. Hence, the DFKE is an ultrafast virtual process in
which no real carriers are created in the sample.  Such nonresonant,
nonperturbative phenomena are observed when the applied field is at the
transition between the classical and quantum regimes, i.e., when the
interaction with the applied field cannot be treated by either neglecting
photon effects or by considering only single photon effects \cite{Chinetal00PRL}.
Recently, Nordstrom {\it et al}.~reported an {\em excitonic} DFKE in
quantum wells using a terahertz free electron laser
\cite{Nordstrometal} while Chin {\it et al}.~observed below band gap absorption
and sideband generation in bulk (optically-thick) GaAs
using intense picosecond MIR pulses \cite{Chinetal01PRL,Chinetal00PRL}.
However, neither the predicted blue-shift of the absorption edge nor the
oscillatory behavior in the spectrum {\em above the band edge} has been
clearly observed in the previous studies.

%Thus $U_p$ varies
%quadratically ($\propto\lambda^2$) with wavelength ($\lambda$). It is this
%fact which leads to unusual power and wavelength dependence of the DFKE as we shall
%see later.  When the photon energy of the radiation is comparable to its
%ponderomotive potential, significant modifications in the electronic states
%occur and cannot be treated perturbatively. The DFKE falls in such a regime
%($U_p \sim \hbar\omega$). As the time-averaged kinetic
%energy of an electron in a driving electric field is equal to $U_p$, a blue
%shift of the band edge by an identical amount ($U_p$) is expected as shown in
%Fig. \ref{cartoon} \cite{JohnsenPRB}.

%%%%%%%%%%%%%% Experimental Setup and Samples %%%%%%%%%%%%%%%%%
We performed a pump-probe study of the transmission near the
band edge of bulk GaAs samples uisng an optical parametric amplifier
(OPA) system with difference-frequency mixing as a
source of tunable MIR pump pulses. The system produced MIR pulses ($\sim$ 150
fs) tunable from 3 to 20 $\mu$m at a 1 kHz repetition rate and was pumped by a
Ti:Sapphire-based chirped pulse regenerative
amplifier (Clark 2010 CPA).  A broadband continuum in the form of white light
was used as the probe beam. The white
light was produced by continuum generation in a sapphire plate, using a small
fraction of the CPA beam.  The broadband probe and the MIR pump beams were then
focused onto the sample using an off-axis parabolic mirror.
A computer-controlled delay stage was used to temporally overlap the two pulses.
After passing through the sample, the spectrum of the probe beam was
dispersed using a grating monochromator and detected using a Si charge-coupled
device camera.

\begin{figure}
\includegraphics [scale=0.6] {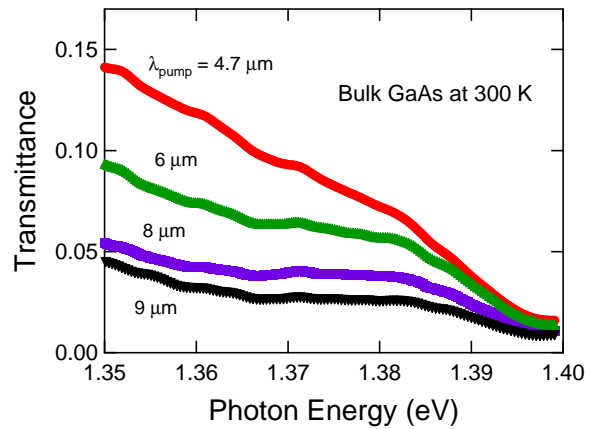}
\caption{Transmission quenching below the band gap in the presence of
strong MIR pump beams of different wavelengths. The effect increases
with increasing wavelength despite the fact that the intensity of our
MIR source decreases for longer wavelengths. This is because the
ponderomotive potential is larger for longer wavelengths in our case.}
\label{quenching}
\end{figure}

%%%%%%%%%%%%%% Induced Below-Bandgap Absorption %%%%%%%%%%%%%%%%%%
We first tried to observe the below band gap absorption predicted by the
theory. For this, a semi-insulating bulk GaAs (band gap $\sim$ 1.42 eV)
wafer, with a thickness of $\sim$ 350 $\mu$m, served as sample.
Shown in Fig.~\ref{quenching}
is the normalized transmission of the probe beam in the presence of the pump
beam below the band edge as a function of photon energy. As can be seen
from the graph, there is a huge absorption of the probe beam below the
band edge which leads to {\em almost complete quenching of the
transmission} (transmission without the pump on the same scale is about 0.5). This
decrease in transmission, due to induced absorption, occurs only during
the temporal overlap of the pump and the probe pulses, confirming the
effect to be a virtual one, i.e., no real carriers and/or lattice-heating
effects are involved.
%The previous observation of below band gap absorption using picosecond
%pulses \cite{Chinetal00PRL} was not as drastic as our observation using
%femtosecond pulses, which is an almost complete quenching of the probe
%transmission.
Furthermore, strikingly, the quenching is more pronounced for longer
wavelengths {\em in spite of the fact that the OPA output intensity
decreased for higher wavelengths}.  This counterintuitive observation can
be explained only by considering the fact that the ponderomotive potential
increases for higher wavelengths due the $1/\omega^2$ dependence
(see Eq.~\ref{Up}).

%%%%%%%%%%%% FIG. 2 %%%%%%%%%%%%%%%%%%%%%%%%%%%

%%%%%%%%%%%%% Absence of Multi-photon Absorption %%%%%%%%%%%%%%
Any multiphoton interband absorption of the pump photons
is ruled out here as this would require the Keldysh parameter $\gamma$
\cite{Keldysh65JETP}, defined as
\begin{eqnarray}
\gamma = \frac{\omega \left( m E_{g}\right)^{1/2}}{eE},
\end{eqnarray}
to be much larger than unity. Here $E_g$ is the band gap and $E$ is the
electric field. In our case, however, we estimate $\gamma$ to
be about 1.8 and hence we are far away from the multiphoton regime.
Also, multiphoton interband absorption would imply generation of
real carriers whose effect would remain even after the pump and probe
pulses do not overlap in time. Thus, we conclude that the observed below band gap
absorption is due to the photon-assisted tunneling of electrons to the
conduction band and can only be attributed to the DFKE.

\begin{figure}
\includegraphics [scale=0.7] {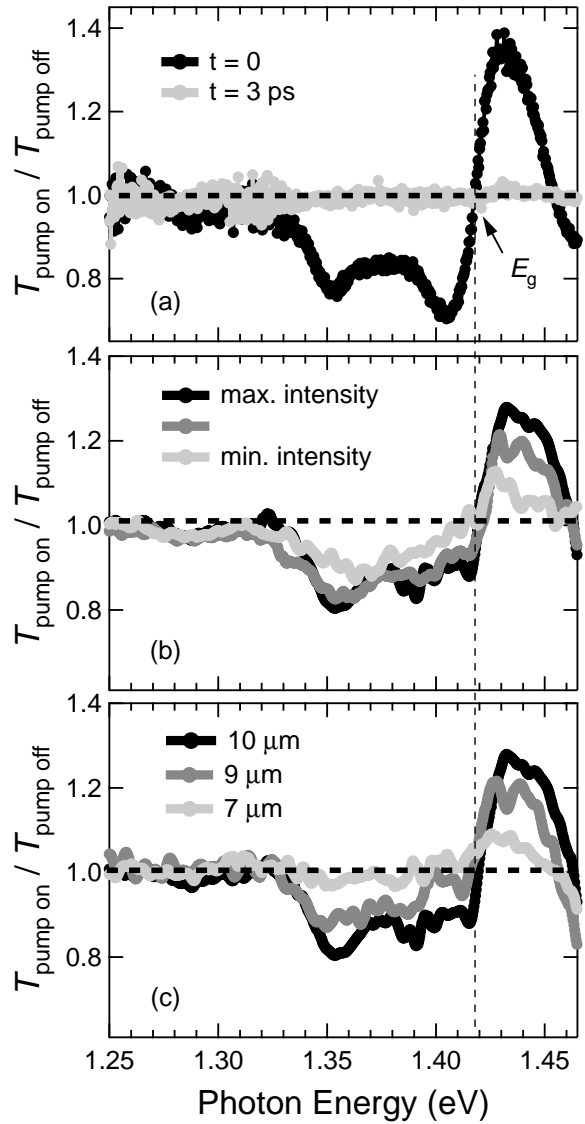}
\caption{(a) Normalized transmission spectra of the white light probe beam
through a GaAs film for zero time delay (black) and a 3 ps time delay
(gray).  The pump wavelength was 9 $\mu$m.
A large ($\sim$ 40 $\%$) induced transparency is observed above the
band gap (1.42 eV) only when the two beams overlap temporally.
(b) The observed change in transmission spectra of the probe beam in the
presence of the pump beam (9 $\mu$m) with different intensities.
(c) Transmission spectra of the probe beam for different pump
wavelengths.  The intensity of the pump beam decreases with increasing
wavelength. The observed effect is more pronounced at longer wavelengths,
even though the intensity of the pump beam decreases with increasing
wavelength.}
\label{transparency}
\end{figure}

%%%%%%%%%%%%%%  Induced Above-Bandgap Transparency  %%%%%%%%%%%%%%%%%%%
In order to observe laser-induced modifications in
transmission spectra above the band edge, a GaAs film sample
($\sim$ 2.3 $\mu$m thick) was used.
Figure \ref{transparency}(a) shows typical data showing induced transparency.
Here the black trace represents the transmission of the white light probe beam
through the film sample in the presence of 9 $\mu$m MIR pump of peak intensity
$\sim$ $10^9$ W/cm$^2$ ($U_p/\hbar\omega \sim$ 0.13) when the pump and probe coincide
temporally; data is normalized to the transmission in the absence of the pump.
A large ($\sim$ 40$\%$) induced transparency is observed right above the
band edge (1.42 eV).
Also shown in Fig.~\ref{transparency}(a) is a gray trace, which was taken
when the pump arrived at the sample 3 ps before the probe and shows
no induced transmission changes.  A detailed time-dependence study
(not shown) indicated that the photoinduced changes (including the transparency)
exist only when the pump and probe pulses overlap temporally.  This fact demonstrates,
once again, the virtual nature of the effect.

%%%%%%%%%%%% FIG. 3 %%%%%%%%%%%%%%%%%%%%%%%%%%%

%%%%%%%%%%%%% Pump-Power Dependence & Wavelength dependence %%%%%%%%%%%%
In Fig.~\ref{transparency}(b), intensity dependence for a fixed pump wavelength
(9 $\mu$m) is shown.  As expected, the photoinduced transparency
decreases for lower intensities of the pump beam.  Finally, the dependence of
the effect on the pump wavelength was studied [Fig.~\ref{transparency}(c)].
The effect is better resolved at longer wavelengths,
a trend similar to the below band gap case.  We did not observe a
significant change in the peak position of the induced transparency for
different wavelengths as we expect $U_p/\hbar\omega$ to be similar in
magnitude ($\sim$ 0.1) for all the cases.  However, a careful examination of
Figs.~\ref{transparency}(b) and \ref{transparency}(c) shows slight shifts of
the peaks when intensity and wavelength of the pump beam are varied.

%%%%%%%%%%%%%%% Simulation Model %%%%%%%%%%%%%%%%%%%%%%%%%%%%%%
The existing models predict an exponential tail (apart from the Urbach tail) in the
absorption spectra below the band gap and oscillations above the band gap.
Also a blue shift of the absorption edge by an amount equal to $U_p$ is predicted.
In order to compare our data with the theory, we use Yacoby's model
\cite{Yacoby68PR}. It calculates the electronic energy levels of a
bulk semiconductor in the presence of a strong pump beam of frequency
$\omega$, which is treated non-perturbatively. The transition rate between
two such states due to an additional perturbation of frequency $\Omega$
(probe beam) is then found. In our experiments, we measured the
time-averaged transmission of the probe beam and hence we calculated the
time-averaged transition rates for valence band to conduction band
transitions using Yacoby's theory [see Eq.~(38) in
\cite{Yacoby68PR}]. Any excitonic contributions were assumed to be negligible
as we are dealing with a bulk semiconductor at room temperature. The values of
intensity, wavelength and effective mass for the calculation were taken to
be $I$ = $10^9$ W/cm$^2$, $\lambda$ = 9 $\mu$m, and $m^*$ = 0.067 $m_e$,
respectively, corresponding to $U_p = 0.13\hbar\omega$.
From the time-averaged change in transition rate, change in
the real and imaginary parts of the dielectric constants were calculated
using Kramers-Kronig relation. Finally, as our sample was thin enough to cause
interference effects by multiple reflections, we calculated the change in
transmission by explicitly taking such effects into account (see, e.g., \cite{Palik}).
Such Fabry-Perot type interference is important mainly below the band gap
where absorption is fairly small.

%%%%%%%%%%%%%%%%%  FIG. 4  %%%%%%%%%%%%%%%%%%%%%
\begin{figure}
\includegraphics [scale=0.65] {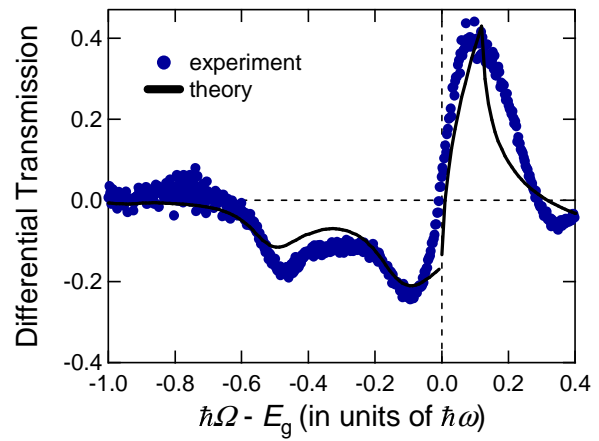}
\caption{Theoretical fit to the experimental data based on Yacoby's theory
(see text).}
\label{fit}
\end{figure}

%%%%%%%%%%%%%%%%%  Comparison  %%%%%%%%%%%%%%%
The calculated and measured differential transmission spectra are plotted in
Fig.~\ref{fit}. For comparison with the theory, the abscissa of the plot was
changed from $\hbar\Omega$ to $\hbar\Omega - E_g$, where $E_g$ is the bandgap of
GaAs, and expressed in units of $\hbar\omega$.
There is good agreement with the theory given the fact that {\em we did
not use any fitting parameters}. Most
importantly, we observe that the large photoinduced transparency is in fact
due the blue-shift of the absorption edge. The magnitude of the blue shift is
equal to our estimate of $U_p$ (0.13$\hbar\omega$) as is expected from
the theory.
However, it should be mentioned
here that we have scaled down the above band gap peak by a factor
of 6. This is justified because in our calculation we have not
included any Urbach-type tail arising from phonons and impurities in the
sample \cite{Urbach1,Urbach2}. Such a tail would actually broaden and
decrease the peak above the band gap.

% By using circularly polarized instead of linearly polarized light,
%spin-selective shift of levels can be envisaged. Such a spin-based effect
%would be similar to the spin AC Stark effect \cite{Guptaetal, Pyroretal}.
%For a shift of 10
%meV (corresponding to $U_p/\hbar\omega$ $\sim$ 0.07 at 9 $\mu$m), this
%would translate to an effective magnetic field of 170 T, assuming a
%g-factor of unity. high magnetic field pulses?

%%%%%%%%%%%%%%%%%% Conclusions %%%%%%%%%%%%%%%%%%%%
In conclusion, we have made the first observation of large
ultrafast induced transparency above the band edge and also huge
absorption below the band edge of GaAs due to the dynamic
Franz-Keldysh effect. We understand this
novel transparency as a result of the blue shift of the band edge. The
existing theory is able to explain our experimental observation reasonably
well and yields an accurate value for the magnitude of the observed blue
shift.

%%%%%%%%%%%%%%%%%% Acknoledgements %%%%%%%%%%%%%%%%%
This work was supported by DARPA MDA972-00-1-0034, NSF DMR-0134058,
and NSF DMR-0325474 (ITR).
 
% BibTeX users please use
%\bibliographystyle{prb}

%\bibliography{ajit}

\end{document}